\documentclass[sigconf,natbib=true,anonymous=false]{acmart}
\usepackage{subcaption}
\usepackage{multirow}
\usepackage{makecell}
\usepackage{listings}

\usepackage{bm}

\AtBeginDocument{%
  \providecommand\BibTeX{{%
    \normalfont B\kern-0.5em{\scshape i\kern-0.25em b}\kern-0.8em\TeX}}}

\copyrightyear{2024}
\acmYear{2024}
\setcopyright{rightsretained}
\acmConference[SIGIR '24]{Proceedings of the 47th International ACM SIGIR Conference on Research and Development in Information Retrieval}{July 14--18, 2024}{Washington, DC, USA}
\acmBooktitle{Proceedings of the 47th International ACM SIGIR Conference on Research and Development in Information Retrieval (SIGIR '24), July 14--18, 2024, Washington, DC, USA}
\acmDOI{10.1145/3626772.3657743}
\acmISBN{979-8-4007-0431-4/24/07}

\makeatletter
\gdef\@copyrightpermission{
  \begin{minipage}{0.3\columnwidth}
   \href{https://creativecommons.org/licenses/by/4.0/}{\includegraphics[width=0.90\textwidth]{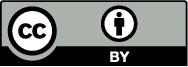}}
  \end{minipage}\hfill
  \begin{minipage}{0.7\columnwidth}
   \href{https://creativecommons.org/licenses/by/4.0/}{This work is licensed under a Creative Commons Attribution International 4.0 License.}
  \end{minipage}
  \vspace{5pt}
}
\makeatother

\begin{document}

\title{Scaling Laws For Dense Retrieval}


\author{Yan Fang}
\authornote{Both authors contributed equally to this research.}
\email{fangy21@mails.tsinghua.edu.cn}
\affiliation{%
  \institution{Department of Computer Science and Technology, Tsinghua University}
  \institution{Quan Cheng Laboratory}
  \state{Beijing 100084}
  \country{China}
}

\author{Jingtao Zhan}
\authornotemark[1]
\email{jingtaozhan@gmail.com}
\affiliation{%
  \institution{Department of Computer Science and Technology, Tsinghua University}
  \institution{Quan Cheng Laboratory}
  \state{Beijing 100084}
  \country{China}
}

\author{Qingyao Ai}
\email{aiqy@tsinghua.edu.cn}
\authornote{Corresponding author}
\affiliation{%
  \institution{Quan Cheng Laboratory}
  \institution{Department of Computer Science and Technology, Tsinghua University}
  \state{Beijing 100084}
  \country{China}
}

\author{Jiaxin Mao}
\email{maojiaxin@gmail.com}
\affiliation{%
  \institution{Gaoling School of Artificial Intelligence, Renmin University of China}
  \state{Beijing 100872}
  \country{China}
}

\author{Weihang Su}
\email{swh22@mails.tsinghua.edu.cn}
\affiliation{%
  \institution{Department of Computer Science and Technology, Tsinghua University}
  \institution{Zhongguancun Laboratory}
  \state{Beijing 100084}
  \country{China}
}

\author{Jia Chen}
\email{chenjia2@xiaohongshu.com}
\affiliation{%
 \institution{Xiaohongshu Inc}
 \state{Beijing}
 \country{China}}

\author{Yiqun Liu}
\email{yiqunliu@tsinghua.edu.cn}
\affiliation{%
  \institution{Department of Computer Science and Technology, Tsinghua University}
  \institution{Zhongguancun Laboratory}
  \state{Beijing 100084}
  \country{China}
}

\renewcommand{\shortauthors}{Yan Fang et al.}

\begin{abstract}

Scaling laws have been observed in a wide range of tasks, particularly in language generation. 
Previous studies have found that the performance of large language models adheres to predictable patterns with respect to the size of models and datasets. 
This helps us design training strategies effectively and efficiently, especially as large-scale training becomes increasingly resource-intensive. 
Yet, in dense retrieval, such scaling law has not been fully explored. 
In this study, we investigate how scaling affects the performance of dense retrieval models. 
We implement dense retrieval models with different numbers of parameters, and train them with various amounts of annotated data. 
We propose to use the contrastive entropy as the evaluation metric, which is continuous compared with discrete ranking metrics and thus can accurately reflect model performance. 
Results indicate that the performance of dense retrieval models follows a precise power-law scaling related to the model size and the number of annotations across different datasets and annotation methods. Additionally, we show that the scaling laws help optimize the training process, such as resolving the resource allocation problem under a budget constraint. We believe that these findings significantly contribute to understanding the scaling effect of dense retrieval models and offer meaningful guidance for future research.

\end{abstract}



\maketitle

\section{Introduction}

\begin{figure*}[!t]
    \centering
    \includegraphics[width=0.8\linewidth]{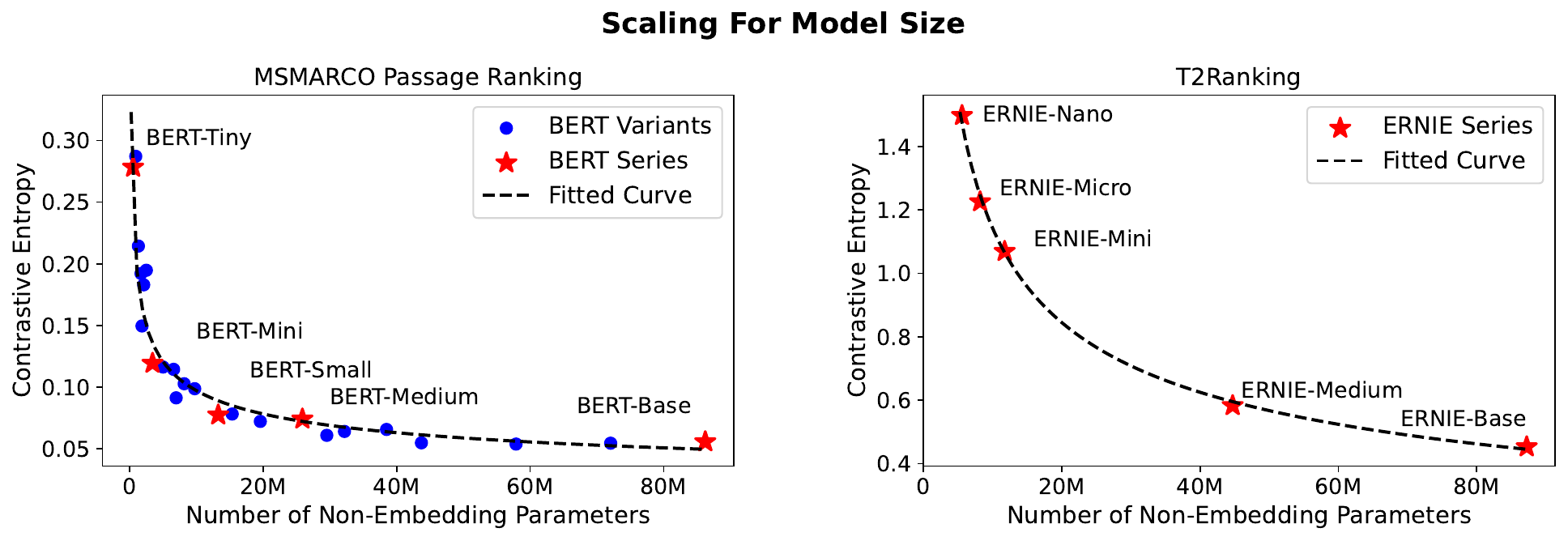}
    \caption{
    Performance of various models on MSMARCO Passage Ranking (left) and T2Ranking (right) datasets. It shows the number of non-embedding parameters (x-axis) and the test-set contrastive entropy (y-axis). The stars and points represent the actual performance. The curves are derived from the scaling law and match the observed data. 
    }
    \label{fig:fig1}
\end{figure*}

The studies of scaling laws in language data can be traced back to a century ago. In the 1920s, a couple of linguisticians discovered that the frequency of a word is proportional to the inverse of its rank when sorting vocabulary based on each word's frequency in the corpus, which is widely known as the Zipf's law~\cite{adamic2002zipf, newman2005power}. Later in the 1960s, Gustav Herdan found that the number of distinct words in a corpus approximately follows a function of the corpus size, which can be approximated with a power function. This is often referred to as the Heaps's law~\cite{lu2010zipf}. These foundational discoveries in scaling laws have profoundly influenced research in linguistics and information retrieval. For example, Zipf's law has inspired the development of several statistical retrieval models, and Heap's law has served as the key principle for the estimation of inverted index, the foundation of many retrieval systems. 

Recently, as language modeling has evolved from statistical analysis to the learning of semantic representations, the focus of scaling law research has also shifted from analyzing text statistics toward the training dynamics of large language models~(LLMs). 
Significant research effort has been dedicated to examining how different factors influence model performance, such as model size, data volume, and computational capacity~\cite{kaplan2020scaling}. 
Findings from these studies reveal precise power law relationships between model performance and scaling factors, which enable researchers and developers to empirically predict model performance without actually constructing the models~\cite{achiam2023gpt}.
Since the training of modern LLMs demands substantial time and financial resources, such scaling laws are of great importance in practice.

Similar to language modeling, Dense Retrieval models have emerged as a significant milestone in this transition from statistical analysis to semantic representation learning in Information Retrieval~\cite{Cai2021SemanticMF,lin2021pretrained}. 
In contrast to traditional statistical retrieval methods such as BM25~\cite{bm25robertson1994some}, Dense Retrieval models are initialized with pre-trained language models and finetuned on annotation data in an end-to-end manner. 
They capture the semantic similarity between queries and documents, and demonstrate superior performance over traditional methods~\cite{qu2020rocketqa,shao2023understanding, ma2023incorporating}. 
However, researchers find that the effectiveness of dense retrieval models is sensitive to multiple training factors~\cite{xiong2020approximateANCE, zhan2021optimizingADORE}. 
Therefore, the construction of effective dense retrieval models under practical constraints (such as budget and latency requirements) is not straightforward, and more insights on the optimization process of Dense Retrieval are needed.

In this paper, we investigate the scaling laws for dense retrieval models\footnote{Code is open-sourced at \url{https://github.com/jingtaozhan/DRScale}.}. 
While some studies have indicated that larger models exhibit improved generalization capabilities in zero-shot dense retrieval tasks~\cite{ni2021large,rosa2022no}, to the best of our knowledge, there isn’t any published literature explicitly discover scaling laws in dense retrieval models.
Specifically, there are two challenges; (1) traditional performance metrics in retrieval tasks (e.g., NDCG) are discrete functions, which limits their ability to stably and smoothly reflect the change of model performance in practice;
(2) the training process of Dense Retrieval involve multiple interrelated factors such as model size, annotation size, and annotation quality, which makes it difficult to isolate the effect of each factor separately. 
To this end, we first propose to evaluate the quality of dense retrieval models with a contrastive entropy metric. 
The idea is inspired by the popular contrastive ranking loss and the analysis of token generation perplexity in LLMs. 
It measures the likelihood of retrieving a relevant document from a randomly sampled candidate set, and shares a similar structure with the training loss of dense retrieval models. 
The smooth nature of this metric considerably facilitates our subsequent analysis. 
Second, to disentangle the effects of model size and data size in dense retrieval, we conducted experiments with models implemented with different pre-trained language models with non-embedding parameter sizes ranging from 0.5 to 87 million, on two of the largest web search datasets, i.e., MSMARCO and T2Ranking. 
Experimental results show that, under proper experimental conditions, the performance of dense retrieval models follows a precise power-law scaling with respect to training factors. Figure~\ref{fig:fig1} illustrates such power-law scaling with model size. 
To investigate the effect of annotation quality, we adopted several LLMs and weak supervision methods to generate training data for dense retrieval models. 
Our results indicate that the observed scaling laws of dense retrieval are uniformly valid across models trained with different types of annotation data. 
Additionally, we show that the joint effect of model and data sizes can be nicely fitted and predicted with a single function within a certain range. 
Such functions can be used to find the best resource allocation strategy given a restricted budget, and could potentially provide important insights for the practical implementation of dense retrieval models and green IR~\cite{strubell-etal-2019-energy}.

This paper is organized as follows. We first breifly revisit the related work in Section~\ref{sec:background}. Then, we present our systematic evaluation framework in Section~\ref{sec:methodology}. With this framework, we investigate the scaling laws of dense retrieval in Section~\ref{sec:scaling_law_dr} and show its potential application in Section~\ref{sec:application}.

\section{Background and Related Work}
\label{sec:background}

In this section, we revisit the background about scaling laws and dense retrieval. We start with the scaling laws in linguistic analysis and in neural language models. Then we present the explorations about dense retrieval techniques and its training technique.

\subsection{Scaling Laws in Linguistic Language Data}
Zipf’s law~\cite{adamic2002zipf, newman2005power} is a well-known evidence about the existence of universal power laws in cognitive science and the social sciences. It shows an inverse correlation between the frequency of a word's occurrence in natural language and its rank in the frequency distribution. It is widely applied in different areas. Furthermore, Zipf's law is tightly connected to other statistical scaling laws in linguistics, notably Heaps' law~\cite{lu2010zipf,gelbukh2001zipf,lansey2009internet}. Heaps' law shows a sublinear growth trajectory between a text's vocabulary size and its total word count. As the total word count increases, the rate of introducing new words diminishes, leading to a plateau in vocabulary expansion. This phenomenon is particularly significant in information retrieval, which serves as the key principle for the estimation of inverted index.

\subsection{Neural Scaling Law}
Neural scaling law describes the relationship between model size, dataset size, computational budget, and performance in neural network training. This concept was first introduced by \citet{hestness2017deep} as a power-law relationship. Subsequently, \citet{kaplan2020scaling} expanded it to larger models. \citet{hoffmann2022an} further refined it by developing a unified formula for scaling laws, incorporating data-dependent scaling terms for compute-optimal training.

These empirical scaling laws offer crucial insights for training large Transformer-based models, particularly by accurately predicting loss. Notably, experimental results from smaller models can be extrapolated to larger ones. Recent studies show that such scaling laws also hold for many other model architectures. For instance, \citet{clark2022unified} investigated the scaling laws in Mixture of Experts (MoE) models. \citet{gao2023scaling} showed the scaling effects in model optimization with Reinforcement Learning. 

Beyond language-centric tasks, these scaling principles have been adapted for domain-specific applications, such as speech recognition~\cite{radford2023robust}, computer vision~\cite{zhai2022scaling,dehghani2023scaling}, and multi-modal language-vision settings~\cite{jia2021scaling,pham2023combined,radford2021learning}. 
In Information Retrieval (IR), \citet{ardalani2022understanding} investigated the application of scaling laws in Click-Through Rate (CTR) recommendation tasks, and \citet{zhang2023scaling} addressed their relevance in conventional ID-based sequential recommendation models. Nonetheless, there has been limited research into whether scaling laws remain applicable in dense retrieval.

\subsection{Dense Retrieval}
We now briefly revisit prior studies in the field of dense retrieval. 
The training data for dense retrieval tasks typically comprises annotated pairs, each consisting of a query and a human-labeled relevant passage. Early research primarily concentrated on effective negative sampling strategies used for dense retrieval training, such as employing random passages or the top irrelevant passages retrieved by BM25 as negative samples~\cite{karpukhin2020dense}. ANCE~\cite{xiong2020approximateANCE} utilized self-mined hard negatives and substantially improved the retrieval performance. Furthermore, \citet{zhan2021optimizingADORE} proposed dynamic hard negatives to further enhance both training efficiency and retrieval effectiveness. RocketQA~\cite{qu2020rocketqa} and TAS-B~\cite{hofstatter2021efficiently} introduced knowledge distillation, utilizing a well-trained cross-encoder model to generate soft labels for training pairs.

Beyond the design of finetuning methods, researchers also explore other techniques, such as pretraining methods and multi-vector retrieval.
(1) Pretraining studies design objectives that are similar to the retrieval tasks. For example, Condenser~\cite{gao2021condenser} and coCondenser~\cite{gao2021unsupervised} use the Sequence Contrastive Learning task to improve the representational capability. RetroMAE~\cite{liu2022retromae} leverages an encoder-decoder architecture, wherein a shallow decoder encourages the encoder to produce higher-quality representations. Contriever~\cite{izacard2022unsupervised} pre-trains dense retrieval models with Inverse Cloze Task and the Independent Cropping Task.
(2) Since the single vector representation in dense retrieval could become a limitation, various studies have explored more complex scoring techniques. ME-BERT~\cite{luan2021sparse} introduces multi-vector representations to enable more precise retrieval of long documents. ColBERT~\cite{khattab2020colbert,santhanam2021colbertv2} investigates token-level vector representations and aggregates scores using a late-interaction mechanism. Other researchers attempt to expand the vector dimension to vocabulary size~\cite{formal2021splade, formal2021spladev2}. This expansion allows dense retrieval models to directly generate term weights, facilitating retrieval similar to sparse models.

Prior explorations of dense retrieval models mainly focus on techniques with a static setup, such as a certain model size, certain data size, etc. Instead, we employ a dynamic setup and explore how model perform when the model size and data size are scaled. 

\subsection{Query Generation}

Besides human-labeled data, dense retrieval can also utilize query generation techniques to generate pseudo annotations~\cite{wang2021gpl, ma2020zero}. Query generation involves generating multiple relevant queries for a given passage~\cite{nogueira2019doc2query, nogueira2019document}. The most basic approach employs unsupervised heuristic methods, such as the previously mentioned Sequence Contrastive Learning (SCL) or Inverse Cloze Task (ICT). However, the quality of the weak supervision data generated by these methods is relatively low. Therefore, they are primarily used in the unsupervised pre-training phase due to their accessibility. More advanced methods leverage pre-trained language models like T5 to generate more precise relevant queries for data augmentation~\cite{nogueira2019doc2query}. Nevertheless, these generated queries are often used for document expansion to enhance the retrieval performance in lexical matching models. As training data, these queries are usually exploited in scenarios where human annotations are scarce, such as in out-of-domain situations.

\section{Methedology}
\label{sec:methodology}

In this section, we first introduce the model architecture and datasets used for exploring the scaling effect of dense retrieval. We further discuss the training strategy used in the experiments and the proposed performance evaluation metrics.

\subsection{Problem Formulation}
We first formalize the dense retrieval model. For a given corpus, the goal is to identify the top relevant passages for a specific query. Dense retrieval models accomplish this by employing an encoder that maps both queries and candidate passages into a shared dense embedding space. Subsequently, a scoring function, such as inner product or cosine similarity, is applied to the encoded dense vectors to compute relevance scores. Let $q$ and $p$ be the query and the passage, respectively. We use $f(\cdot; \theta)$ to denote the the mapping function of the dense retrieval model parameterized by $\theta$. The relevance score $s(q, p)$ is as follows:
\begin{equation}
    s(q, p) = \left \langle f(q;\theta), f(p;\theta)\right \rangle 
\end{equation}
In this paper, we only consider that the encoders for queries and passages are shared, as it is a popular implementation choice in practice. We leave the studies of separate query and document encoders to future studies.

The training data for dense retrieval typically comprises a set of training queries and associated human annotations. Each query is annotated with one or more relevant passages, and the remaining unannotated passages are generally presumed irrelevant. In this paper, we adhere to this annotation standard and consider each query-positive-passage pair as an individual data point. Formally, the training set  consists of $n$ data points, $\{(q_i, p_i^+)\}_{i=1}^n$, where $q_i$ and $p_i^+$ denote the $i$-th query in the training set and its corresponding annotated positive passage.

\subsection{Model Architechture}

With the development of large-scale pre-trained language models, advanced dense retrieval models in recent years have followed the Transformer's structure. While some studies have explored using decoder-only architectures to generate dense vector representations of texts, mainstream dense retrieval models still employ encoder-only models such as BERT due to its bi-directional modeling ability. Formally, a pre-trained Transformer, augmented with a projection layer, serves as the text encoder:
\begin{equation}
\label{eq:encoding}
    v = \left( {\rm Transformer} (x)\right) W + b
\end{equation}
where $x$ represents the text input, and $W$ and $b$ are the parameters of the projection layer.

Typically, the generated vector representation is derived from the [CLS] token representation (in BERT series models) or the mean pooling of the outputs from the last Transformer layer. The main function of the projection layer is to map these vectors into the target semantic space.

In our study, we experimented with Transformer models of various model sizes. With limited annotated query-passage pairs, it is usually difficult to train a large dense retrieval model from scratch. As a result, most dense retrieval models are initialized with pre-trained language models and then perform fine-tuning on the annotated data. Therefore, to align with prevailing research practices, we focus our analysis on dense retrieval models initialized from different sizes of pre-trained language models. 

Previous studies have shown that different pre-training tasks significantly affect the performance of dense retrieval models~\cite{gao2021condenser,gao2021unsupervised,izacard2022unsupervised,liu2022retromae,maxinyu}.
To minimize such influence, we select a series of models with identical pre-training configurations and only differ in parameter sizes. Specifically, for experiments on the English corpus, we chose 24 BERT checkpoints from the original Google release~\cite{devlin2018bert}, with model sizes ranging from 0.5 million (BERT-Tiny) to 82 million parameters (BERT-Base)\footnote{https://github.com/google-research/bert. Following \citet{kaplan2020scaling}, we define the model size as the number of non-embedding parameters.}. For experiments on Chinese retrieval benchmarks, we selected the ERNIE series~\cite{sun2019ernie}, which were pre-trained on Chinese corpora using tasks similar to BERT. To each model, we attach a projection layer, as shown in Eq.~(\ref{eq:encoding}), to map the output dimensionality of embeddings to 768 for consistent comparisons.

\subsection{Training Data}
We utilize publicly available retrieval datasets for exploring the scaling effect for dense retrieval models. To ensure the generalizability and completeness of our study, we follow recent DR research and use MS MARCO Passage Ranking dataset~\cite{nguyen2016msmarco} (English) and T2Ranking~\cite{xie2023t2ranking} (Chinese) for the experiments. MS MARCO Passage Ranking is a large-scale annotated dataset with a corpus of 8.8M passages from English web pages and 0.5M training queries. Each training query is coupled with a manually labeled positive passage, which together constitute the annotated pairs. MS MARCO also provides around 7,000 validation queries for performance evaluation. T2Ranking is a recently released large-scale Chinese benchmark for passage ranking, which comprises more than 300k queries and over 2M unique passages collected from real-world search engines. 

\subsection{Training Setting}
As discussed previously, in this paper, we construct dense retrieval models from the pre-trained language model checkpoints and perform fine-tuning with the annotated query-document pairs in each dataset. One of the most important parts of dense retrieval model training is the negative sampling strategy. Previous work has shown that mining hard negative samples in the training process can significantly improve the retrieval performance. However, the primary objective of this work is to investigate the scaling effects of dense retrieval models. As a result, we do not focus on sophisticated training strategies. For simplicity, we adopt the most straightforward approaches, namely random negative sampling and in-batch negative techniques, for the training of all dense retrieval models in this paper. These methods are employed to minimize the influence of sampling strategies.

Formally, for each query-passage pair ($q_i, p_i^+$), we randomly select a set of unlabeled passages from the corpus as the negative. Then we can optimize the following contrastive ranking loss:
\begin{align}
    \mathcal{L}(\theta) 
    &=-{1\over B}\sum_{i=1}^B \log{\exp\left(s(q_i, p_{i}^+;\theta)\right) \over {\exp\left(s(q_i, p_{i}^+;\theta)\right) + \sum_{j} \exp\left(s(q_i, p_{j}^-;\theta)\right)}}
\end{align}
where $B$ denotes the training batch size, $\{p_j^-\}$ is the set of negative passages and $s(q, p;\theta)$ is the scoring function of query and passage:
\begin{align}
    s(q, d; \theta) 
    &= \left \langle f(q;\theta), f(d;\theta)\right \rangle
\end{align}
Here, $\left \langle \cdot \right \rangle$ denotes inner product and $\theta$ denotes the parameters of the text encoder. 

We fine-tune the models for a fixed 10,000 steps and random sample 256 negatives at each step.

\begin{figure*}[!h]
    \centering
    \includegraphics[width=0.9\linewidth]{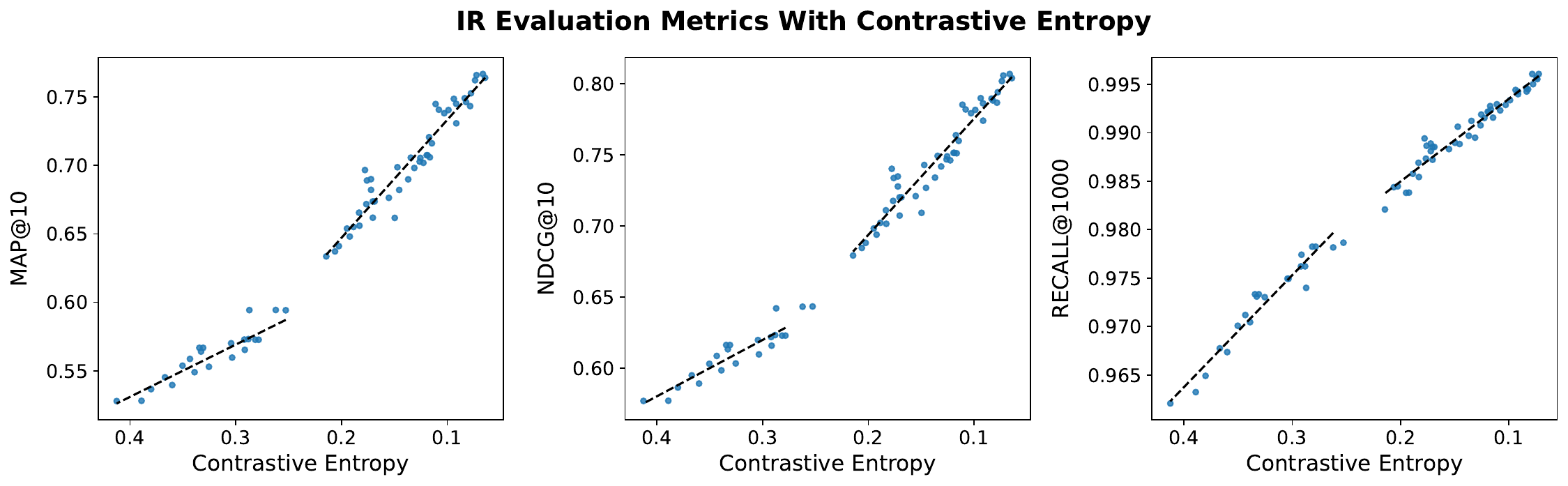}
    \caption{
    Relationship between standard ranking metrics and contrastive entropy for different Dense Retrieval models on the MSMARCO Passage Ranking dataset. The figures illustrate the contrastive entropy (x-axis) versus standard ranking metrics (y-axis). The results indicate a strong positive correlation. Besides, the figures highlight an emergent ability phenomenon~\cite{wei2022emergent} around a contrastive entropy value of approximately 0.25, where there is a significant improvement in ranking metrics.
    }
    \label{fig:fig9}
\end{figure*}

\subsection{Evaluation Protocol}

We now discuss how we evaluate the retrieval performance. The most widely adopted retrieval paradigm is to rank passages in the corpus based on the relevance scores predicted by the retrieval model and retrieve the Top-K candidates to form a ranked list. The performance of the retrieval model is then assessed based on the ranked list using well-defined ranking metrics such as NDCG@K and MAP@K. However, such metrics are not continuous due to their discrete nature and reliance on a cutoff parameter, K. Because the ranking metrics of a ranked list would not change unless the sequence of the passages changes, these ranking metrics are not sensitive to the changes of model outputs in many cases. Also, with the cutoff in ranking metric, a positive passage only contributes to the metric when ranked within the top K results. If it falls beyond K, whether at K+1 or further, it has no impact on the metric score. The characteristics of these existing ranking metrics make them unsuitable for the investigation of scaling laws in dense retrieval. 

To solve these problems, we propose to utilize a continuous metric that sensitively reflects the overall retrieval capability of the models. Inspired by the analysis of scaling laws in large language models, which utilize the perplexity of token generations as evaluation metrics, we propose to use the contrastive entropy as our evaluation metric. Formally, for each query-passage pair in the test set, we randomly select a fixed number (256 in this paper) of negative passages and define the contrastive entropy as:
\begin{align}
    -\log{\exp\left(s(q_i, p_{i}^+;\theta)\right) \over {\exp\left(s(q_i, p_{i}^+;\theta)\right) + \sum_{j} \exp\left(s(q_i, p_{j}^-;\theta)\right)}}
\end{align}

We investiagte the correlation between the contrastive entropy and existing ranking metrics. We train multiple dense retrieval models. To efficiently evaluate their retrieval performance, we sample a subset corpus that contains 100,000 passages during evaluation. Figure~\ref{fig:fig9} shows the contrastive entropy and ranking metrics, including MAP@10, NDCG@10, and Recall@1000. We can see that the correlation between the contrastive entropy and existing ranking metrics is strong and positive. It is close to a linear correlation. Therefore, we believe that using contrastive entropy is an effective measure to assess the overall retrieval ability of models in our study.

Figure~\ref{fig:fig9} also shows a critical point around $0.25$ contrastive entropy, where the top ranking performance evaluated with traditional metrics substantially improves. We attribute this phenomenon to emergent ranking ability. Concurrently, \citet{du2024understanding} also observe this phenomenon in generation tasks. They find emergent abilities are tightly related to a certain loss value. We leave further exploration to future studies.

\begin{figure*}[!t]
    \centering
    \includegraphics[width=0.8\linewidth]{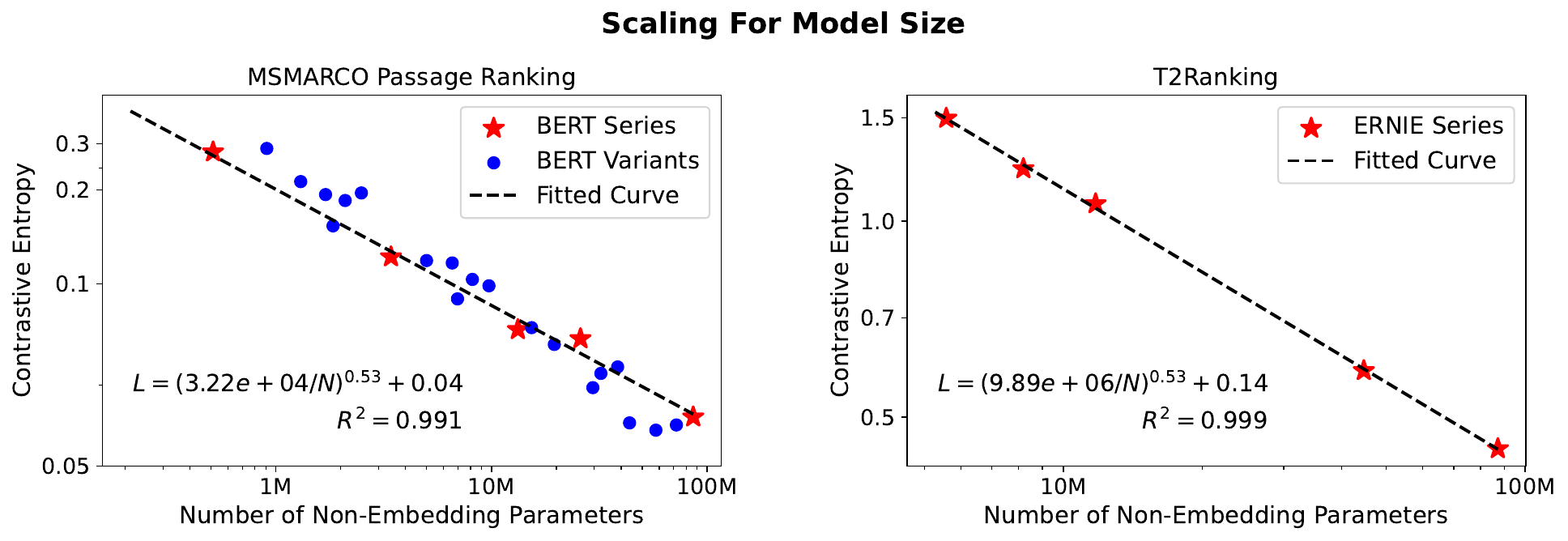}
    \caption{
    Scaling laws for model performance as a function of model size on MSMARCO Passage Ranking (left) and T2Ranking (right) datasets. The figures display the contrastive entropy (y-axis) against the number of non-embedding parameters (x-axis, logarithmic scale) for different models. Points and stars represent the actual performance, aligning closely along a straight line. The dashed lines are fitted using Eq.~(\ref{eq:model_size_scaling_law}), demonstrating a close match with the empirical data.
    }
    \label{fig:fig1_log}
\end{figure*}

\section{Scaling Laws For Dense Retrieval}
\label{sec:scaling_law_dr}

In this section, we show the results of our experiments and summarize our initial investigation of the scaling laws for dense retrieval. Specifically, we aim to thoroughly investigate the following three research questions:
\begin{itemize}
\item How does model size impact dense retrieval performance?
\item How does annotated training data size influence dense retrieval performance?
\item Do different types of data annotations result in distinct scaling effects on dense retrieval models?
\end{itemize}

\subsection{Model Size Scaling}
We finetune models of various sizes using the human-annotated training pairs. The finetuning is performed on the entire training sets. We do not utilize early stopping and instead report the best test set loss throughout the training process. This is mainly to mitigate the influence of suboptimal early stopping, which could lead to models being underfitted or overfitted.

Figure~\ref{fig:fig1_log} illustrates the contrastive entropy on the test set with respect to model sizes. As shown in the figure, the retrieval performance improves (indicated by a lower test loss) as the model size increases. On the left side of the diagram, red stars represent the official checkpoints of variously sized BERT models, while blue points denote other official variants released concurrently. These variants differ in aspects such as the number of attention heads or feed-forward dimensions. The right diagram, in contrast, only features red stars, as the different shape variants of ERNIE are not publicly available.

Based on the observation, we propose to fit the scaling law in terms of model sizes as follows:
\begin{equation}
\label{eq:model_size_scaling_law}
    L(N) = \left( {A \over N} \right)^{\alpha} + \delta_N
\end{equation}
where $N$ represents the number of non-embedding parameters of the model, and $L(N)$ denotes the model's contrastive entropy on the test set. Parameters $A$, $\alpha$ and $\delta_N$ are the coefficients.

Note that we introduce a parameter $\delta_N$, which represents a irreducible loss term. It means that a sufficiently large model (setting $N$ to infinity) can only reduces the loss to $\delta_N$ rather than zero. This irreducible loss is reasonable given the incomplete annotations and subjective understanding of relevance. On one hand, some relevant passages may not be annotated because they are not sucessfully recalled and are outside the annotation pool. On the other hand, relevance may be subjective to different annotators, which results in even imperfect agreement among different human annotators. Consequently, it is hard for models to perfectly agree with human annotations. Therefore, we believe there should be a irreducible term in the scaling law.

We employ least squares method to fit the linear curve. The coefficients are detailed in Table~\ref{tab:model_size_param}. The coefficient of determination (R²) suggests a good fit. Based on these results, we validate that the contrastive entropy follows a power-law scaling in relation to the size of non-embedding parameters. 

\begin{figure*}[!t]
    \centering
    \includegraphics[width=0.8\linewidth]{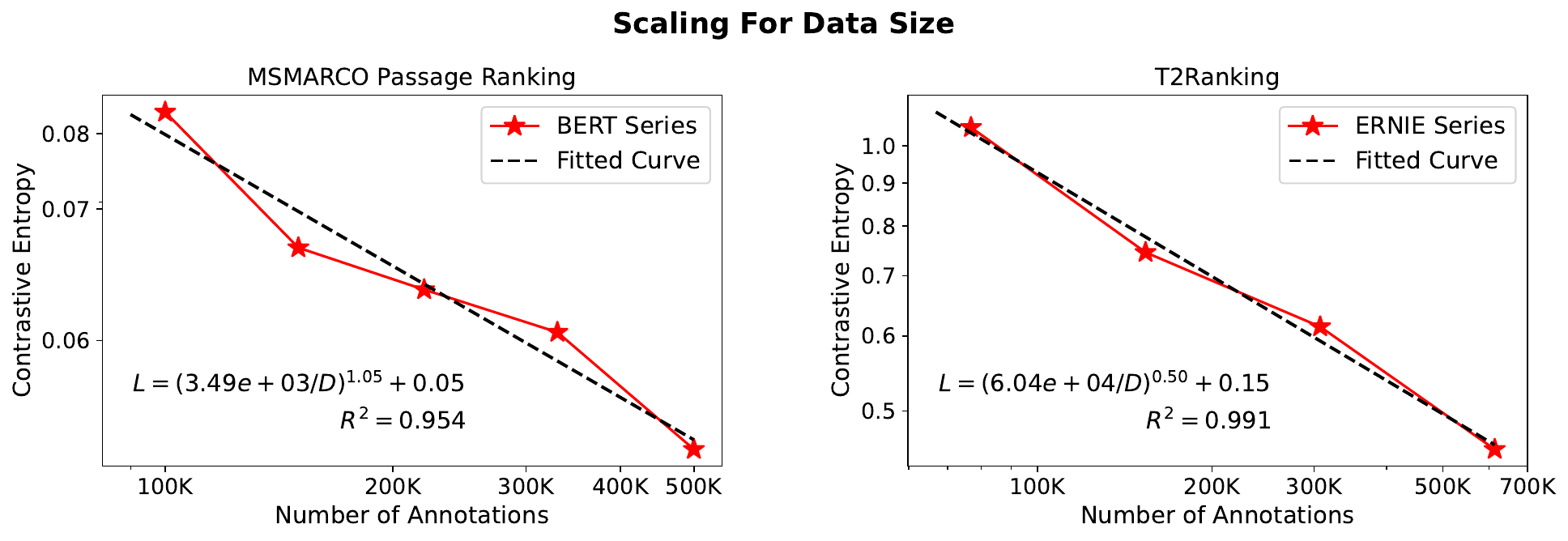}
    \caption{
    Scaling laws for model performance relative to training data size on MSMARCO Passage Ranking (left) and T2Ranking (right) datasets. The figures illustrate the contrastive entropy (y-axis) as a function of the number of annotated query-passage pairs (x-axis, logarithmic scale) for a fixed model size. Points and stars show the actual performance, aligning closely with a straight line. The dashed lines are fitted using Eq.~(\ref{eq:data_size_scaling_law}), demonstrating a strong fit with the empirical data.
    }
    \label{fig:fig3}
\end{figure*}

\begin{table}[!tbp]
\caption{Fitting parameters for model size scaling}
\label{tab:model_size_param}
\begin{tabular}{ccccc}
\hline
Dataset  & $A$ & $ \alpha $ & $\delta_N$ &$R^2$ \\ 
\hline
MSMARCO   & $3.22 \times 10^4 $   & 0.53 &   0.04 & 0.991 \\ 
T2Ranking   & $9.89 \times 10^6 $   & 0.53 &   0.14 & 0.999  \\
\hline
\end{tabular}
\end{table}

Such discoveries offer new perspectives for future research experiments. For example, given this scaling law, we can initially train smaller models, fit the corresponding scaling curves, and then extrapolate them to predict the performance of larger models. This significantly reduces the cost of conducting experiments directly on larger models and instead offers the opportunity to experiment with different training strategies on smaller models to validate the effectiveness of new approaches.

\subsection{Data Size Scaling}
We then fix the model size and vary the size of the training data, defined by the number of annotated query-passage pairs. To minimize potential underfit problems caused by small models, we finetune the largest model in this experiment, i.e., the BERT-Base model. Here we present the experiment results up to using all available annotation data. 

The results are shown in Figure~\ref{fig:fig3}. Similarly, we fit the scaling law in terms of data size with the following log-linear curve:
\begin{equation}
\label{eq:data_size_scaling_law}
    L(D) = \left( {B \over D} \right)^{\beta} + \delta_D
\end{equation}
where $D$ represents the number of annotated query-passage pairs, and $L(D)$ denotes the contrastive entropy. $B$, $\beta$ and $\delta_D$ are coefficient to be estimated. The coefficient of determination (R²) indicates a good fit. Based on these results, we infer that the contrastive entropy follows a power-law scaling relative to the number of annotated query-passage pairs, with specific parameters detailed in Table~\ref{tab:data_size_param}.

\begin{table}[!t]
\caption{Fitting parameters for data size scaling}
\label{tab:data_size_param}
\begin{tabular}{ccccc}
\hline
Dataset  & $B$ & $ \beta $ & $\delta_D$ &$R^2$ \\ 
\hline
MSMARCO   & $3.49 \times 10^3 $   & 1.05 &   0.05 & 0.954 \\ 
T2Ranking   & $6.04 \times 10^4 $   & 0.50 &   0.15 & 0.991  \\
\hline
\end{tabular}
\end{table}

This finding offers an alternative perspective for future annotation process. For instance, to determine the amount of annotations for a new corpus, the traditional approach relies on past experience without a clear understanding of the sufficiency of data annotation. With the data-size scaling law, a potential approach is initiating with a minimal amount of annotations, training a model, and fitting the corresponding scaling curve. Accordingly, we can approximate the necessary size of data annotation based on the target performance of the dense retrieval model. This approach establishes a clear relationship between data annotation and the desired performance outcomes. It allows researchers to have a precise expectation of future model performance, facilitating more effective planning and budgeting for annotation tasks.

\subsection{Annotation Quality}
So far, we have observed strong scaling phenomena of dense retrieval model performance with respect to model sizes and data sizes. Yet, in the IR scenario, another aspect that remained unexplored is the quality of data annotations: \textit{Does the scaling effect hold true for data of different quality?} 

To investigate this, we conduct experiments using annotations of different quality. Due to constraints in time and resources, our experiments are exclusively conducted on the MSMARCO Passage Ranking dataset. We employ query generation techniques to create three distinct types of annotations:
\begin{itemize}
	\item \textbf{Inverse Cloze Task (ICT):} ICT extracts sentences from passages and and uses the sentence as pseudo-query for the passage. Since it ignores the semantic information, the generated data is of low quality.
	\item \textbf{Supervised Generation Models:} We utilize docT5query~\cite{nogueira2019doc2query} to produce multiple queries for each passage. DocT5query is trained on human annotations. The generated data is of higher-quality than ICT's.
	\item \textbf{Large Language Models (LLMs):} We instruct LLMs to generate relevant queries for given passages. Since LLMs are strong in language understanding and generation, we consider the data quality to be better than both ICT and docT5query. We adopt ChatGLM3~\cite{zeng2023glmb} due to its impressive performance in various tasks. The prompt for query generation is shown in Appendix~\ref{sec:appendix}.
\end{itemize}

\begin{figure}[!t]
    \centering
    \includegraphics[width=0.85\linewidth]{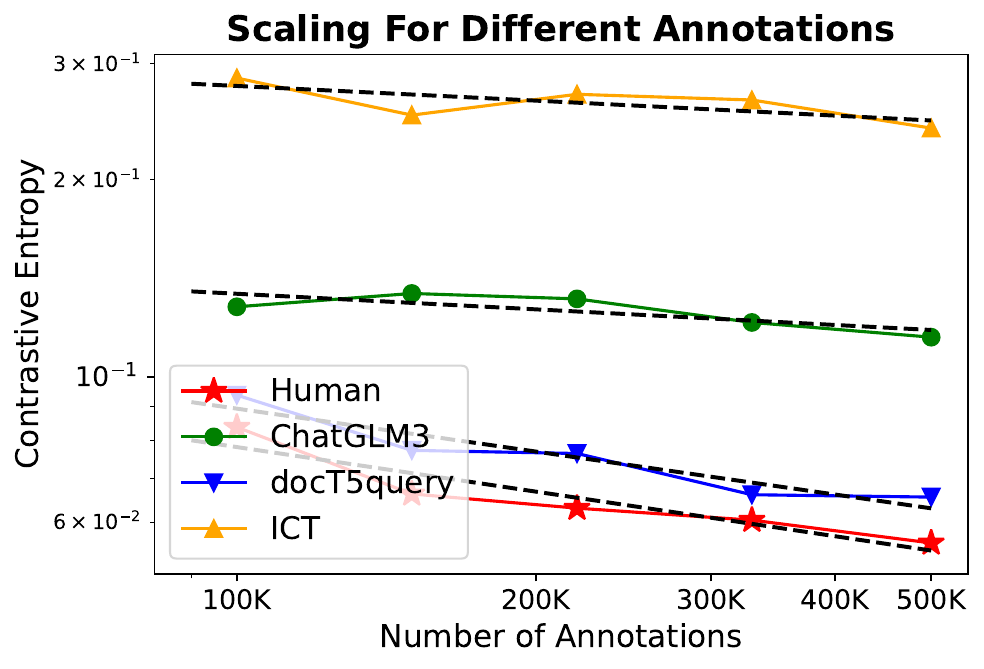}
    \caption{
    Scaling effects of annotation quality for retrieval performance on MS MARCO. Dashed lines are fitted using Eq.~(\ref{eq:data_size_scaling_law}), which demonstrate the power-law scaling across different annotation methods. ChatGLM3 annotations exhibit the steepest slope and surpass human annotations at 500k.    
    }
    \label{fig:fig4}
\end{figure}

For ICT and ChatGLM3, we generate a query for each positive document annotated by humans in the original datasets. For docT5query, we randomly sampled 500,000 passages from the corpus for query generation, since it is originally trained on the human annotated passages. In this way, we can align the training passages with human annotations and other annotations. Also, it's important to note that, despite employing different data generation methods, our evaluations consistently utilize the human-annotated development set. The results are reported in Figure~\ref{fig:fig4}.

We can see that the retrieval performance scales with respect to different annotation qualities. Comparing the three methods of query generation, the log-linear curve of ICT exhibits the smallest slope. This observation aligns with our expectation that ICT is a weak supervision method and limits the enhancements for retrieval models when we increase the data size. The data quality from ChatGLM3 is better, but not as good as docT5query. This is because that docT5query has been finetuned on this dataset while ChatGLM3 is used in a zero-shot manner. Moreover, we use a 6B ChatGLM3 instead of a very large model, which may also result in its sub-optimal performance.
Among all these methods, human annotations lead to the best-performing models. Therefore, there is still a large room for improvment about using large language models to generate pseudo training data. 


\subsection{Model-Data Joint Laws}

\begin{figure}[!t]
    \centering
    \includegraphics[width=0.85\linewidth]{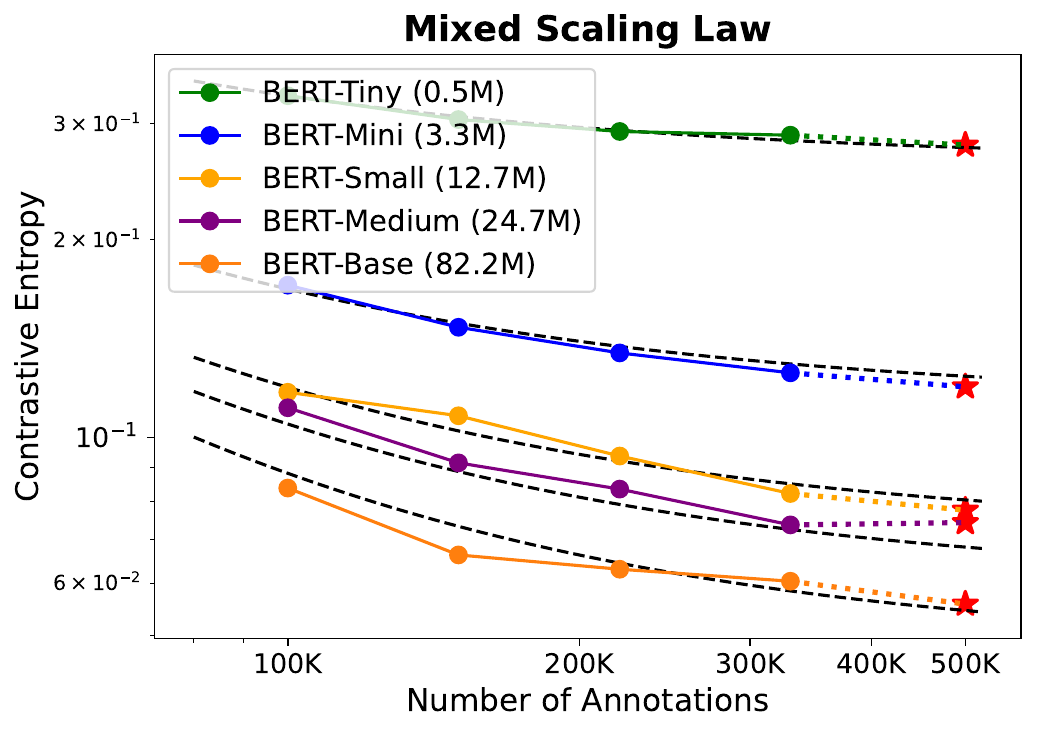}
    \caption{
    Modeling the joint effects of model size and data size on retrieval performance using a unified scaling law. 
    Solid dots are used for fitting, while the red stars are performance to predict. The dashed lines are fitted with Eq.~(\ref{eq:joint_law}) and closely aligns with the observed data. 
    }
    \label{fig:fig10}
\end{figure}

We combine the above observations into a single function that can characterize the joint effects of model size and data size. Inspired by the scaling laws of LLMs~\cite{kaplan2020scaling}, we employ the following equation to describe the scaling effect:
\begin{align}
\label{eq:joint_law}
    &L(N, D) = \left[ \left( {A \over N}\right)^{\alpha \over \beta} + {B \over D}  \right]^{\beta} + \delta \\
    &A \approx 3.6 \times 10^4,~~B \approx 7.1 \times 10^3 \\
    &\alpha \approx 0.56, ~~\beta \approx 1.31,~~ \delta \approx 0.03
\end{align}
where $N, D$ represents the model size and data size, respectively, and $A, B, \alpha, \beta, \delta$ are coefficients. 
We employ results with different model sizes and data sizes to estimate the coefficients. Figure~\ref{fig:fig10} illustrates the actual contrastive entropy and the predictions. In this figure, solid dots represent the data used for curve fitting, while the dashed line indicates the resulting fitted curve. The red stars denote data points utilized to evaluate the accuracy of our predictions. We can see that the predictions relatively are close to the real values.

\section{Application in Budget Allocation}
\label{sec:application}

\begin{figure}[!t]
    \centering
    \includegraphics[width=0.8\linewidth]{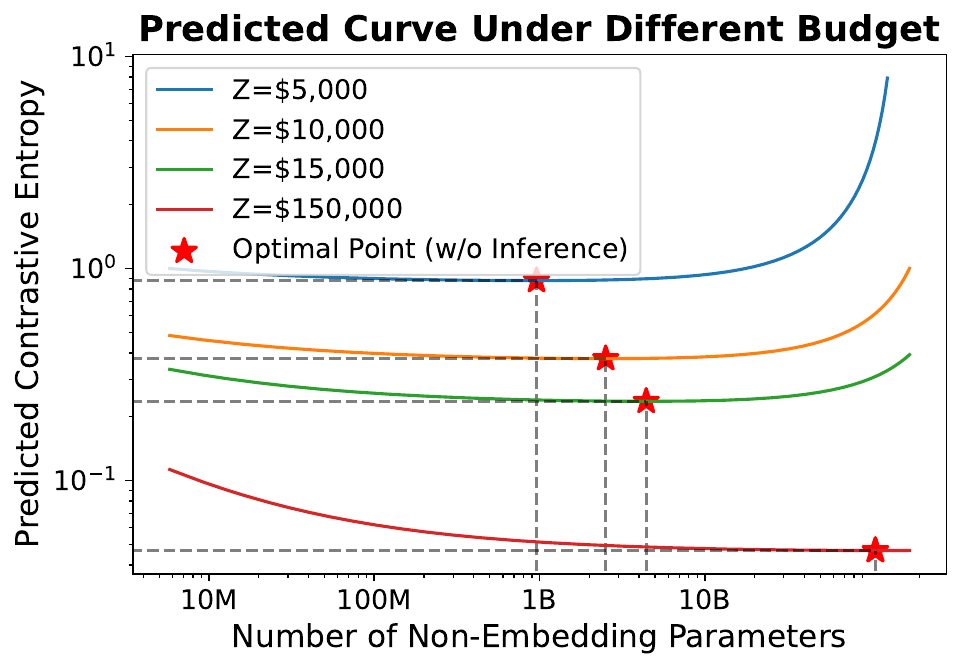}
    \caption{
    Predicted contrastive entropy for different model sizes under varying cost budgets, excluding inference costs. 
    With an increase in model size, performance initially improves due to higher data efficiency of larger models, but eventually degenerates because of limited data annotation.
    }
    \label{fig:fig7}
\end{figure}

In this section, we showcase a potential application of the scaling laws for dense retrieval observed in our experiments. We use Eq.~\ref{eq:joint_law} in this section. 

We attempt to estimate the comprehensive cost associated with the lifecycle of dense retrieval models, including data annotation, model training, and model inference. The total cost of training a model with $N$ parameters using $D$ data points is given by:
\begin{equation}
    Z(N, D) = Z_{\rm data} \cdot D + Z_{\rm train} \cdot N + Z_{\rm infer} \cdot N
\end{equation}
Here, $Z_{\rm data}, Z_{\rm train}, Z_{\rm infer}$ represent cost factors corresponding to annotations, training, and inference, respectively. 

Now we estimate the approximate values for $Z_{\rm data}, Z_{\rm train}, Z_{\rm infer}$. The cost of human annotations is approximated at \$0.6 per query-passage pair~\cite{althammer2023annotating}. For computational costs, according to previous studies~\cite{kaplan2020scaling,clark2020electra}, the training and inference computation for Transformer can be assumed by $6N$ and $2N$ FLOPs, respectively. We refer to common cloud computing and the price for using an A100 80G GPU is assumed to be \$3.93 per hour\footnote{\url{https://cloud.google.com/compute/gpus-pricing}}, with the peak computational power around 312 TFLOPs. For the training phase, we assume that the model is trained for 10,000 steps on a single A100 GPU. At each step, the model encodes a query, a positive passage and a negative passage with a batch size of 256. Each query is around 30 tokens and each passage is around 60 tokens. For the inference phase, we assume that the model is employed in a web search engine. Based on public statistics, we assume that there are around 30 trillion web pages in Google's index\footnote{From \url{https://en.wikipedia.org/wiki/Google\_Search}, the estimated size of Google's index is around 30 trillion in 2012.}. The inference cost for a dense retrieval model predominantly involves encoding the entire corpus. We estimate that each web page contains approximately 512 tokens. 
We assume the GPU utilization efficiency is 25\%, then we have
\begin{align}
    Z_{\rm data} &\approx 0.6 \\
    Z_{\rm train} &\approx {10000 \times (30 + 2 \times 60) \times 256 \times 6 \times 3.93  \over  312T \times 3600 \times 25\%} = 3.22 \times 10^{-8}\\
    Z_{\rm infer} &\approx {30 \times 10^{12} \times 512  \times 2 \times 3.93  \over 312T * 3600 \times 25\%} = 0.43 
\end{align}

\begin{figure}[!t]
    \centering
    \includegraphics[width=0.8\linewidth]{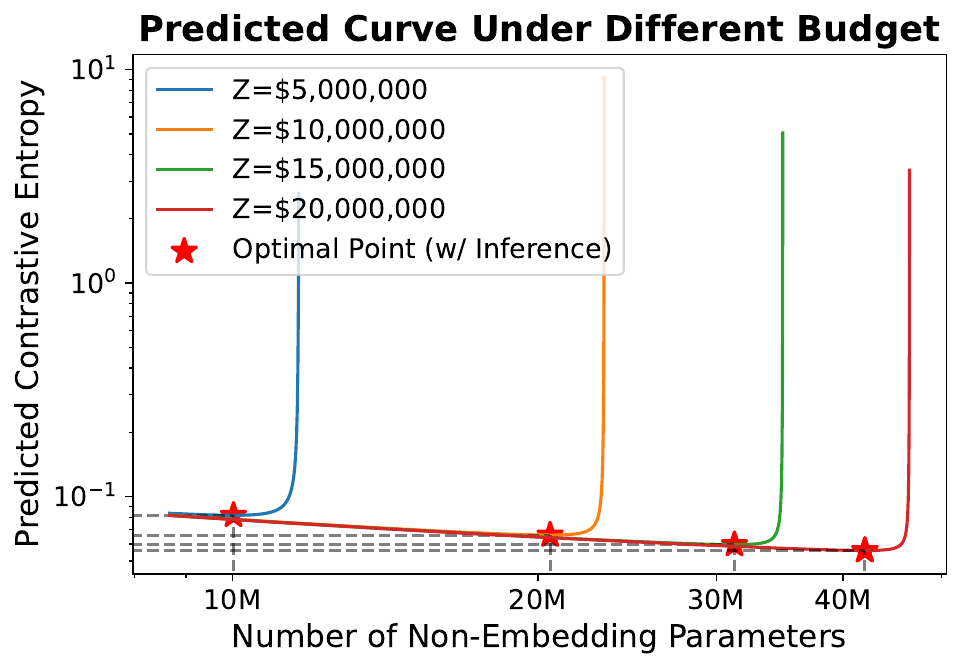}
    \caption{
    Predicted contrastive entropy for different model sizes under varying cost budgets, including inference costs. 
    Inference is much more costly than training and results in small models be the optimum.
    }
    \label{fig:fig8}
\end{figure}

We first excludes the cost of inference and only focuses on annotation and training. Figure~\ref{fig:fig7} shows the predicted contrastive entropy against model size under different cost budget. It is clear that for a fixed cost budget, as the training model size increases, the predicted retrieval performance initially improves and then slowly degenerates. The improvement is because that a relatively larger model is more data-efficient and can exhibit stronger ranking performance than smaller models. Nevertheless, when models are too large, only limited budget can be used for data annotation. The data limitations make the performance degenerate. Overall, if we do not consider inference cost, for a 20,000\$ budget, it is optimal to train a model with 13 billion parameters. This is primarily due to that larger models are more data-efficient and that human annotation is significantly more expensive than training. Therefore, under a limited budget, maximizing model size can yield better results.

We further include the inference costs into this analysis. The result is shown in Figure \ref{fig:fig8}. It is clear that the optimal model size significantly decreases to only million-scale parameters, even under a larger budget. This is because that the inference cost is huge compared to training cost ($Z_{\rm infer} \gg Z_{\rm train}$) and that the small models are more inference-efficient. A billion-scale model will make the inference cost prohibitively high.

\section{Limitation and Future Work}

This study pioneers the investigation of scaling laws in dense retrieval. We cover major factors like model scale, datasets, training volume, and annotation methods. Several other aspects remain unexplored and can be addressed in future research.

Our experiments utilize contrastive entropy as the evaluation metric due to its continuity, which addresses the discrete nature of ranking metrics and facilitates the derivation of scaling laws. Although we demonstrate a positive correlation between contrastive entropy and ranking performance, it is important to note that they are not equivalent. For instance, as shown in Figure~\ref{fig:fig9}, similar contrastive entropy scores do not guarantee similar ranking performance. Future research may explore alternative metrics that might offer a more direct correlation with ranking outcomes.

The training process in this work is based on random negative sampling and contrastive learning. We do not cover more sophisticated training techniques, such as hard negative sampling~\cite{xiong2020approximateANCE, zhan2021optimizingADORE}, distillation~\cite{hofstatter2021efficiently}, and contrastive pre-training~\cite{izacard2022unsupervised, gao2021unsupervised, gao2021condenser}. These methods could potentially influence the scaling behaviors observed and should be investigated in the future.

We focus on a common dense retrieval architecture where text is mapped to a single dense vector of fixed dimensionality. However, some researchers have experimented with variations in this architecture, such as mapping to vectors of varying dimensions~\cite{reimers-gurevych-2021-curse}, multiple vectors~\cite{luan2021sparse, khattab2020colbert}, or even sparse vectors~\cite{formal2021splade, mallia2021learning}. Future work could explore how these architectural modifications impact the scaling laws for dense retrieval.

Our evaluations are conducted within in-domain datasets. Although we also attempt out-of-domain test (not reported in the paper), the available datasets are relatively small and yield unstable results, making it challenging to draw robust conclusions. Thus, our results do not currently account for out-of-domain scenarios, and more extensive evaluation could be beneficial in future work.

While we try to assess scaling across various scales, our resources limit the maximum size of our models and the extent of data size. Future work could further evaluate scaling laws on an even larger scale with more extensive models and annotations.

\section{Conclusion}

This paper systematically investigates the scaling laws of dense retrieval. We conduct experiments on both Chinese and English datasets to assess the impact of model size, data size, and annotation methods on retrieval performance. By utilizing contrastive entropy as the metric, we observe a power law relationship between performance and both model size and data size across different annotation methods and datasets. We also show that the scaling laws help optimize training processes. For instance, the scaling laws is important to budget allocation management, as demonstrated in our experiments. Moreover, scaling laws allows to evaluate the efficacy of different annotation methods. As shown in our experiments, there is still a large improvement room for using large language models to generate relevance annotations. We believe scaling laws offer a systematic approach to assess and improve the training processes of ranking models. While this study has laid a foundation for future exploration in this area, further research is needed to expand our understanding of scaling laws across more varied domains, scales, architectures, and evaluations.

\appendix

\section{Appendix}
\label{sec:appendix}
We use the following prompt for ChatGLM3 to generate queries for one passage. Note that \{\} is the placeholder for the actual passage.
\begin{lstlisting}[caption={ChatGLM3 Prompt for Query Generation.}, label={lst:code_example}]
Please generate 5 relevant queries according to the given passage for search purpose.
1. Each query should be relevant to the passage.
2. Each query should be around 10 to 20 words.
3. Please generate diverse queries.
4. Output in JSON format, with keys: "query1", "query2", "query3", "query4", "query5".
5. Please respond in English. DO NOT use Chinese.
Passage: {}
\end{lstlisting}

\begin{acks}
This work is supported by Quan Cheng Laboratory (Grant No. QCLZD202301).
\end{acks}

\bibliographystyle{ACM-Reference-Format}
\bibliography{references}

\end{document}